\documentclass[apj]{emulateapj}

\usepackage{mathptmx}
\usepackage{color}
\usepackage{subfigure}
\usepackage{bm}

\journalinfo{Draft version \today}
\slugcomment{ApJ submitted}
\shorttitle{The Mass of the Black Hole in Nova Muscae 1991}
\shortauthors{WU ET AL.}

\def\simgt{\lower 2pt \hbox{$\, \buildrel {\scriptstyle >}\over {\scriptstyle \sim}\,$}}
\def\simlt{\lower 2pt \hbox{$\, \buildrel {\scriptstyle <}\over {\scriptstyle \sim}\,$}}

\def\xray{\hbox{X-ray}}

\def\nmus{NovaMus}

\defcitealias{wu+2015b}{Paper~I}

\begin{document}


\title{The Mass of the Black Hole in the X-ray Binary Nova Muscae
  1991}



\author{\rm Jianfeng~Wu\altaffilmark{1},
Jerome~A.~Orosz\altaffilmark{2},
Jeffrey~E.~McClintock\altaffilmark{1},
Imran~Hasan\altaffilmark{3},
Charles~D.~Bailyn\altaffilmark{3,4},
Lijun~Gou\altaffilmark{5,6},
Zihan~Chen\altaffilmark{5,6}
}


\altaffiltext{1}
     {Harvard-Smithsonian Center for Astrophysics, 60 Garden Street,
     Cambridge, MA 02138, USA} 
\altaffiltext{2}
     {Department of Astronomy, San Diego State University, 5500
     Campanile Drive, San Diego, CA 92182, USA} 
\altaffiltext{3}
     {Department of Astronomy, Yale University, P.~O.~Box 208101, New
     Haven, CT 06520, USA}
\altaffiltext{4}
     {Yale-NUS College, 6 College Avenue East, Singapore, 138614}
\altaffiltext{5}
     {National Astronomical Observatories, Chinese Academy of Sciences,
     Beijing 100012, China} 
\altaffiltext{6}
     {University of Chinese Academy of Sciences, Beijing 100012, China} 


\email{jianfeng.wu@cfa.harvard.edu}


\begin{abstract}

  The optical counterpart of the black-hole soft X-ray transient Nova
  Muscae 1991 has brightened by $\Delta{V}\approx0.8$ mag since
  its return to quiescence 23 years ago. We present the first clear
  evidence that the brightening of soft X-ray transients in quiescence
  occurs at a nearly linear rate. This discovery, and our precise
  determination of the disk component of emission obtained using our
  {\it simultaneous} photometric and spectroscopic data, have allowed us
  to identify and accurately model archival ellipsoidal light curves of
  the highest quality.  The simultaneity, and the strong constraint it
  provides on the component of disk emission, is a key element of our
  work. Based on our analysis of the light curves, and our earlier
  measurements of the mass function and mass ratio, we have obtained for
  Nova Muscae 1991 the first accurate estimates of its systemic
  inclination $i=43.2^{+2.1}_{-2.7}$ deg, and black hole mass 
  $M=11.0^{+2.1}_{-1.4}\ M_\odot$.  Based on our determination of the
  radius of the secondary, we estimate the distance to be
  $D=4.95^{+0.69}_{-0.65}$~kpc.  We discuss the implications of our work
  for future dynamical studies of black-hole soft \xray\ transients.

\end{abstract}

\keywords{black hole physics --- stars: black holes --- binaries:
  general --- X-rays: binaries}



\section{Introduction}\label{intro}

Stellar-mass black holes are identified and studied in \xray\ binary
systems in the Milky Way and nearby galaxies \citep{remillard+2006}. The
compact objects in two dozen of these systems are dynamically confirmed
to be black holes with masses in the range $M=5$--30$~M_\odot$ \citep[for
the most recent review, see][]{casares+2014}. One remarkable property of
the mass distribution of these black holes is the ``mass gap'', i.e.,
the lack of black holes with $M=3-5~M_\odot$
\citep[e.g.,][]{ozel+2010,farr+2011}.  Assuming that this mass gap is
not caused by selection biases \citep[e.g.,][]{narayan+2005}, it
provides an important constraint on supernova models
\citep{fryer+2001,belczynski+2012,kochanek+2014}.  

Mass measurements, along with estimates of distance $D$ and systemic
inclination $i$, have also made possible demonstrably reliable
measurements of the spins of ten black holes via the continuum-fitting
method (\citealt{mcclintock+2014}, and references therein).  In turn,
these spin measurements benefit a variety of astrophysical
studies. For example, they are the basis for a correlation between jet
power and black hole spin
(\citealt{narayan+2012,steiner+2013,mcclintock+2014}; but also see 
\citealt{fender+2010,russell+2013}), which provides insights into the
energy generation mechanism of relativistic jets \citep{narayan+2014}.
In testing this correlation and its associated jet model, it is
essential to increase the sample size, which is presently only five
black-hole soft X-ray transients (which we refer to hereafter as
BHSXTs). Especially important for the correlation is a single
short-period BHSXT, A0620-00 ($P=7.8$~hr), which anchors the correlation
at low spin and low jet power. In this paper, we report estimates of
$M$, $i$ and $D$ for Nova Muscae 1991 (GS/GRS 1124$-$683; hereafter
\nmus), another short-period BHSXT ($P=10.4$~hr), which is very similar
in many respects to A0620-00 \citep{remillard+1992}. The compact primary
of \nmus\ was dynamically confirmed to be a black hole shortly after its
discovery by \citet{remillard+1992}.  Using our values of $M$, $i$ and $D$
for \nmus, we will go on to estimate the black hole's spin (Chen et
al. 2015) and test the jet model of \citet{narayan+2012} by
comparing our spin estimate to the value predicted in
\citet{steiner+2013}. 

In measuring the mass of a black hole in an \xray\ binary, one must
determine three parameters: 1) the value of the mass function $f(M)$,
which sets a hard lower limit on the mass of the compact object; 2) the
ratio of the mass of the secondary star to that of the black hole, $q
\equiv M_2/M$; and 3) the orbital inclination angle $i$ of the 
system. The first two parameters are usually obtained via spectroscopy
in the optical or near-infrared (NIR) band.  The value of the mass
function is determined by the semi-amplitude of the radial velocity
curve of the secondary $K_2$ and the orbital period $P$:
\begin{equation}
f(M) \equiv \frac{{P}K_{2}^{3}}{2\pi G} = \frac{M\sin^3i}{(1+q)^{2}}.
\end{equation}
It is relatively straightforward to determine $q$ by measuring the
rotational broadening of the photospheric lines $v\sin i$
\citep[e.g.,][]{wu+2015b}. Usually, the greatest challenge is obtaining
an accurate estimate of $i$, since none of the BHSXTs discovered to date
exhibits an X-ray eclipse, as discussed by \citet{narayan+2005}.

While estimates of $i$ have been obtained for three BHSXTs by modeling
jet data \citep[e.g.][]{fender+1999,steiner+2012b,steiner+2012a}, the
common approach to constraining $i$ is to model multi-color optical/NIR
light curves while the system is in \xray\ quiescence.  Ideally,
during each orbital cycle the ellipsoidal light curve of the
tidally-distorted secondary, which fills its Roche lobe, shows two
equal maxima and two unequal minima. A near-ideal example of such a
light curve is that of GRO~J1655$-$40 for which several groups have
obtained estimates of $i$ that are in good agreement
\citep{orosz+1997,vanderhooft+1998,greene+2001,beer+2002}. In fact,
GRO~J1655$-$40 is the only BHSXT for which the error in the black hole
mass is dominated by the uncertainties in the mass function and the mass
ratio, rather than the uncertainty in $i$ \citep{casares+2014}.

For most BHSXTs, however, the quiescent ellipsoidal optical/NIR light
curves are contaminated by a non-stellar contribution from the accretion
disk (and possibly the jet). This component, which can vary rapidly,
often distorts the light curve, and it sometimes completely conceals the
ellipsoidal modulation. This ``disk-veiling'' issue is most problematic
for the short-period systems (defined here as $P<$12~hrs) which
generally have relatively faint $K$- or $M$-type
secondaries. \citet{kreidberg+2012} discuss three types of non-stellar
optical/NIR emission. One is a constant ``pedestal'' level of emission
from the disk that dilutes the amplitude of the ellipsoidal modulation.
The second is a component that varies periodically with the orbital
period of the system, producing asymmetric distortions in the light
curves, thereby increasing or decreasing the amplitude of the
ellipsoidal-component of modulation. We model this component as
emission from a wedge-shaped hotspot on the disk. These two disk
components of light can be included in the light-curve model.

In addition, there is a third and more problematic component, namely,
aperiodic flickering due to variability in the accretion
flow. \citet{cantrell+2008} elucidated this component by identifying two
principal distinct optical state for A0620$-$00 in quiescence, the {\it
  passive} and {\it active} states.  Cantrell et al.\ concluded that
only light curves in the passive state, which have minimal aperiodic
variability, are suitable for modeling in order to estimate $i$; those
in the active state are dominated by aperiodic flickering and are
unsuitable.  Cantrell et al.\ also found that A0620$-$00 in the active
state is brighter than in the passive state by by $\sim0.3$ mag,
which is expected as the accretion disk ``builds up'' during quiescence
\citep{lasota+2001}.  \citet{kreidberg+2012} further investigated the
potential systematic uncertainties of using active-state data in
estimating $i$ and $M$.

In summary, accurate measurements of $i$ and $M$ for BHSXTs requires
that one selects and models light curves obtained in the passive quiescent
state. Even in this case, careful modeling of the pedestal and hotspot
components is required; this is particularly true in the case of the
short-period systems, such as A0620--00 and \nmus.  However, in most
earlier studies of BHSXTs, the disk contribution was only crudely
estimated, or it was ignored altogether, while no distinctions were drawn
between passive- and active-state data.  Consequently, in a number of
cases, there has been wide disagreement in the inclination estimates
that have been obtained for the same system \citep[see summary in
Table~1 of][]{casares+2014}.

Earlier studies of \nmus\ have provided estimates of $f(M)$, $q$, $i$,
and $M$ \citep[e.g.,][]{orosz+1996,casares+1997,gelino+2001}.  However,
in no case has the disk contribution to the optical/NIR light curves
been robustly constrained. In the most recent work, \citet{gelino+2001}
obtained an estimate of $i=54^\circ \pm1.5^\circ$ by modeling the $J$
and $K$ band light curves assuming that the disk contribution is
negligible, an assumption that was subsequently shown to be ill-founded
\citep{reynolds+2008,kreidberg+2012}. Based on the mass measurement of
Gelino et~al., \citet{morningstar+2014}
found a retrograde spin ($a_*=-0.25^{+0.05}_{-0.64}$) for the black hole
in \nmus, which is a remarkable result for a black hole in a BHSXT,
given that the spin of the black hole is believed to be accrued
gradually over the lifetime of the system via accretion torques
\citep{fragos+2015}.

In this work, we present the first accurate measurements of the
inclination of the binary system and the mass of the black hole, as well
as an estimate of distance. The accuracy of our results derives from (i)
the unprecedented quality of our determination of the disk veiling and
the quality of our radial velocity data; (ii) use of the first reliable
measurement of the mass ratio $q$; and (iii) the exclusive use of
passive light-curve data in modeling the ellipsoidal variability,
namely, data obtained shortly after the system returned to quiescence
following its 1991 outburst \citep{orosz+1996}. In our earlier paper
on the dynamics of \nmus\ \citep[hereafter Paper~I]{wu+2015b}, we 
extensively discuss items (i) and (ii).  Meantime, item (iii) is a
feature topic of this work (\S\ref{lc:pick}).

As elaborated in \citetalias{wu+2015b}, in 2009 we tackled the problem
of the variable 
component of non-stellar emission by making phase-resolved spectroscopic
and photometric observations of \nmus\ that were strictly simultaneous
($<1$~s).  The spectroscopic and photometric data were both collected at
Las Campanas Observatory using the Magellan/Clay 6.5~m and du~Pont 2.5~m
telescopes, respectively.  We obtained 72 high-resolution spectra using
the Magellan Echellette spectrograph (MagE; \citealt{marshall+2008}),
which cover the wavelength range 3000--10000~\AA.  Based on independent
measurements for several echellette orders, we obtained the following
precise and robust results: $K_2=406.8\pm2.7$~km~s$^{-1}$ and $v\sin
i=85.0\pm2.6$~km~s$^{-1}$, which respectively imply $f(M)=3.02\pm0.06\
M_\odot$ and $q=0.079\pm0.007$.  Of paramount importance for this work
in modeling the light curves, we obtained a precise constraint on the
fraction of the light due to the disk in 2009: $56.7\pm1.4\%$ in the
$V$-band. 

In this work, we use the results described above from
\citetalias{wu+2015b} to model 
the multi-color light curves of \nmus\ in order to obtain reliable
estimates of $M$, $i$, $D$, and other parameters for \nmus. The structure
of this paper is as follows. In \S\ref{lc}, we first document the steady
brightening of \nmus\ during quiescence. Then, from all available data,
we select the light curves that are suitable to model, and for each
light curve we constrain the fraction of the light contributed by the
disk. In \S\ref{model} we detail the procedures we use to model the
data, and we present the final adopted values of $M$ and $i$, and all
other model parameters.  In \S\ref{dist} we estimate the distance to
\nmus, a parameter that is crucial for measuring the spin via the
continuum-fitting method. Our summary and discussion are presented in
\S\ref{discuss}.


\section{Photometric Data}\label{lc}

\subsection{Optical Brightening of \nmus\ during X-ray
  Quiescence}\label{lc:smarts}

The $V$-band light curve we obtained in 2009 contains a strong and
variable, non-stellar component of light \citepalias{wu+2015b}, and it is
grossly inferior for ellipsoidal modeling compared to the passive-state
light curves obtained in 1992--1993 by \citet{orosz+1996}.  Furthermore,
the system was 0.65 mag brighter in $V$ in 2009 than it was in
1992--1993.  These results motivated us to investigate how \nmus\
brightened during this period, whether regularly and gradually, or
chaotically.

\begin{figure*}[t]
    \centering
    \includegraphics[width=5in,angle=90]{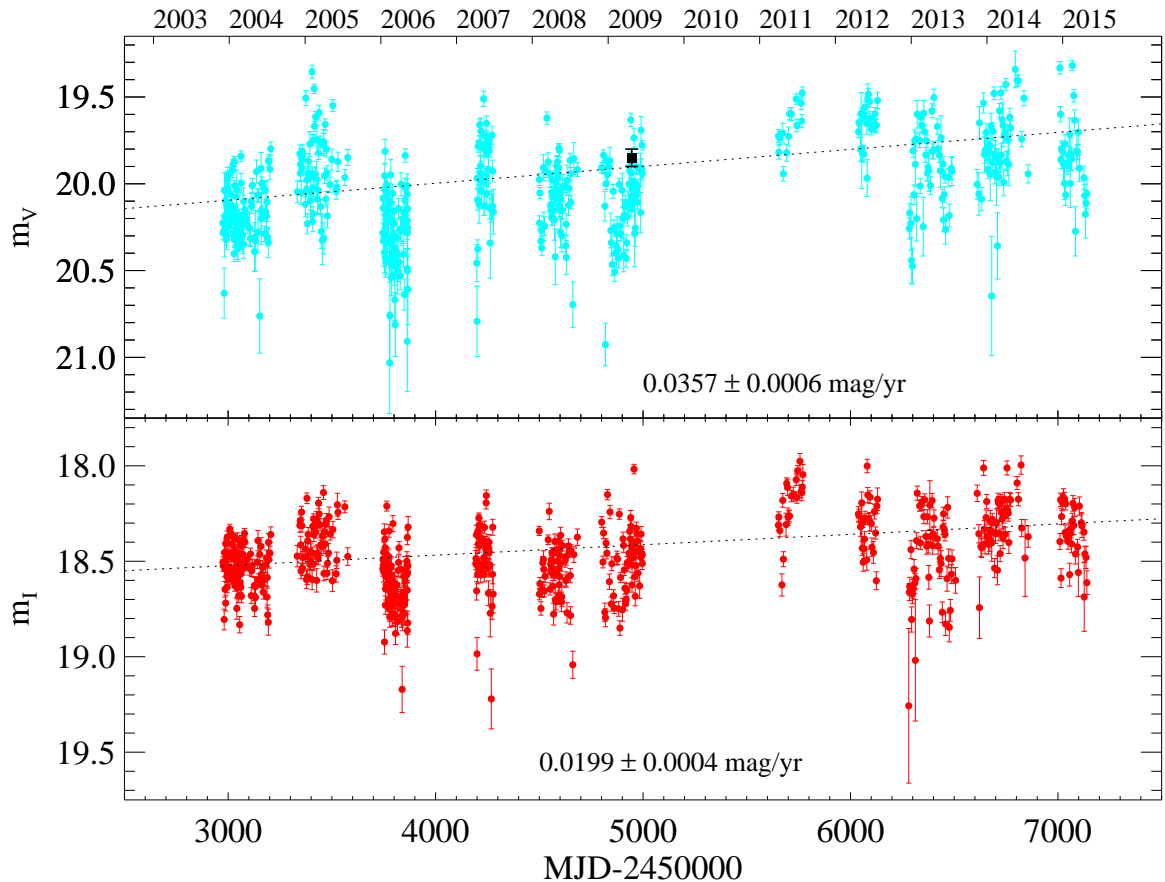}
    \caption{\footnotesize{SMARTS light curves of \nmus\ for the period
        2003--2015 in the $V$-band (upper panel; filled cyan circles)
        and in the $I$-band (lower panel; filled red circles). The
        dashed line and rate of brightening given in each panel are the
        result of a linear fit to the data.  The filled black square
        represents our mean $V$-band magnitude of \nmus\ obtained in
        2009 with the du~Pont telescope.}
    \label{smarts_fig}}
\end{figure*}%

Long-term photometric monitoring of \nmus\ in \xray\ quiescence has
been performed on a near-nightly basis using the Small and Moderate
Aperture Research Telescope System
(SMARTS)\footnote{\url{http://www.astro.yale.edu/smarts.}}.  We made
use of the $V$- and $I$-band photometric data for the years
2003--2015, sans 2010. The data set comprises 556 frames in the $V$
band and 592 frames in the $I$ band (Fig.~\ref{smarts_fig}).  We
performed differential photometry using nearby, calibrated 
standard stars with $V-I$ colors similar to those of \nmus.  Our
estimates of photometric error (shown in Fig.~\ref{smarts_fig}), which are
statistical, are typically $\sim0.05$~mag for the $V$-band and
$\sim0.04$~mag for the $I$-band.  Our estimate of the uncertainty in
the zero point for each band is $\sim0.05$~mag.

Fig.~\ref{smarts_fig} shows for both bands a clear trend of brightening
of \nmus\ over the course of 12 years. Applying a least-square linear
regression to all the data points, we find that \nmus\ brightened at a
rate of $0.0357\pm0.0006$~mag~yr$^{-1}$ in the $V$-band and
$0.0199\pm0.0004$~mag~yr$^{-1}$ in the $I$-band.  {\it This is the first
  direct evidence of a gradual and a near-linear brightening of a BHSXT
  during X-ray quiescence.} It is worth noting that the rate of
brightening is greater in the $V$-band than the $I$-band; i.e., the
pedestal of disk emission becomes bluer as the system brightens. This
may be caused by an increase in the temperature of the disk, or by a
change in the structure of the disk (e.g., a decrease in the inner-disk
radius).

The brightening is predicted by the disk instability model \citep[DIM;
for a review see][]{lasota+2001}.  According to the DIM, and as these
light curves show, after its return to quiescence following an outburst
the disk of a BHSXT slowly but steadily builds via accretion, and the
system brightens as the pedestal component grows \citep[see Fig.~3
of][]{dubus+2001}.  Meanwhile, the DIM does not predict the growth of
aperiodic flickering and the distortions that develop in the ellipsoidal
light curves as the quiescent system transitions from the passive
optical state to the active state \citep{cantrell+2008}. According to
the DIM, eventually a disk instability is triggered and a new outburst
begins when the surface density of the disk at some radius reaches a
critical value.  The outburst recurrence time for BHSXTs can be years or
decades. \nmus\ has remained in quiescence for 24 years, ever since its
only known outburst in 1991.  By including disk irradiation and
truncation into the instability model, while modeling the inner
accretion flow as an advection dominated accretion flow
\citep[ADAF;][]{narayan+1994}, \citet{dubus+2001} reproduced the long
recurrence times for BHSXTs.

\subsection{Selection of Light Curves}\label{lc:pick}

As discussed above, only light curves obtained in the passive state are
useful in attempting to constrain the inclination. Unfortunately, the
high-quality $V$-band data we obtained in 2009 using the du~Pont
telescope are unsuitable because \nmus\ was in an active state, as
evidenced by strong aperiodic flickering \citepalias[see Fig.~8
of][]{wu+2015b}.  Likewise, the SMARTS light curves are dominated by
aperiodic variations, showing that \nmus\ has been in an active state
since at least 2003. Therefore, the du~Pont and SMARTS light curves
cannot be expected to provide reliable constraints on the inclination.

We therefore turned to examine the highest-quality archival light curves
in the literature, namely, those of \citet{orosz+1996} and
\citet{gelino+2001}, while passing over the $H$-band light curve of
\citet{shahbaz+1997} because of its less desirable quality. We first
discuss the light curves of \citet{orosz+1996}.  They were obtained in
1992--1995, shortly after the 1991 outburst of \nmus, and they are thus
minimally contaminated by disk emission. The authors present light
curves in two bands: a $B+V$-band and a wide $I$-band, which have
central wavelengths of $\approx5000$~\AA\ and $\approx9000$~\AA,
respectively.  Orosz et al.\ report that in 1992 \nmus\ had a $V$-band
magnitude of $20.51\pm0.07$, which was fainter than we observed in 2009
by $0.66\pm0.09$ mag.  Based on these two data points, the average
rate of brightening for the 17-year period 1992--2009 is
$0.0388\pm0.0053$~mag~yr$^{-1}$, which is consistent with the rate
of $0.0357\pm0.0006$~mag~yr$^{-1}$ determined using SMARTS data for
2003--2015 (\S\ref{lc:smarts}).

\begin{figure}[t]
    \centering
    \includegraphics[width=3.4in]{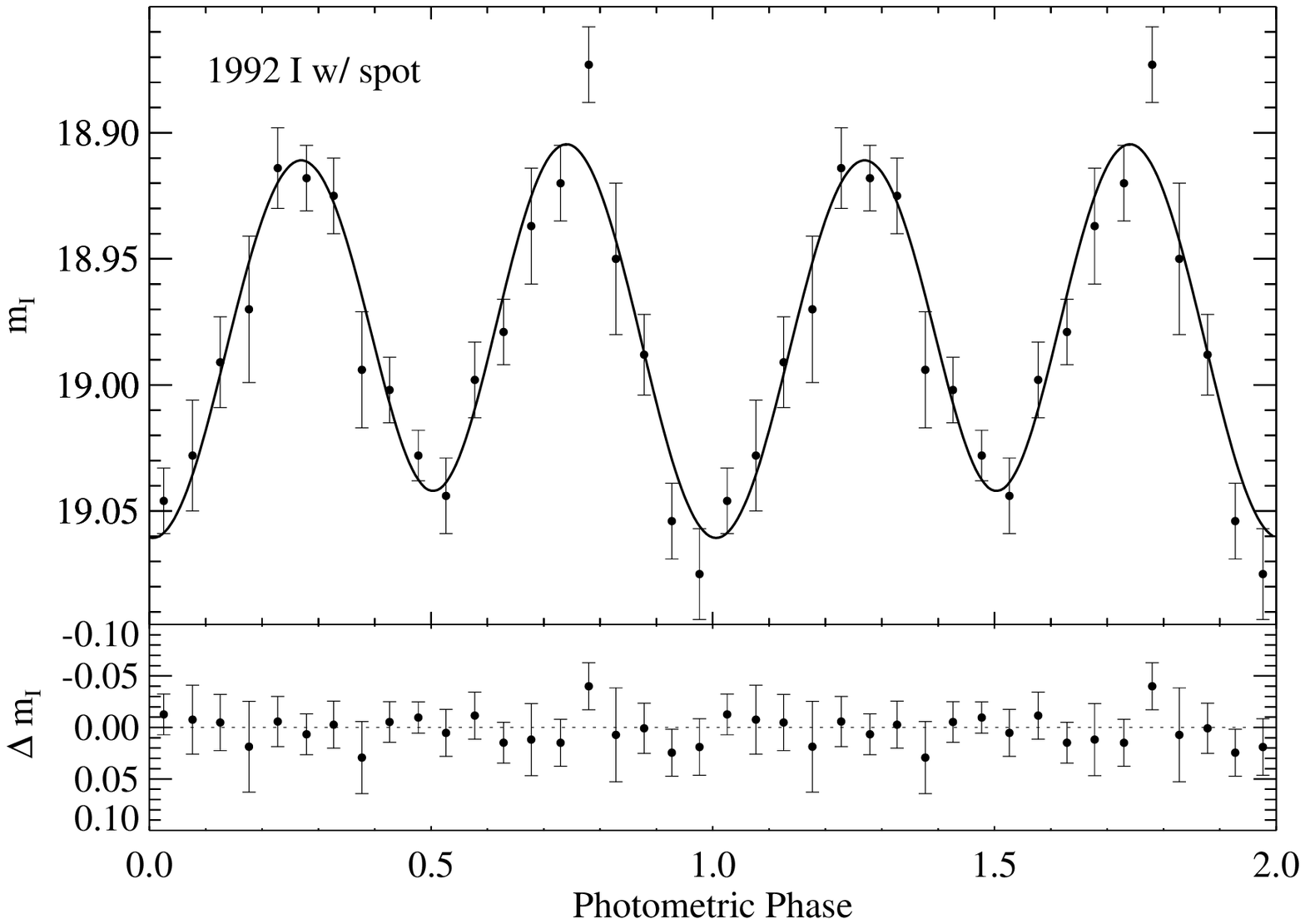}\\
    \includegraphics[width=3.4in]{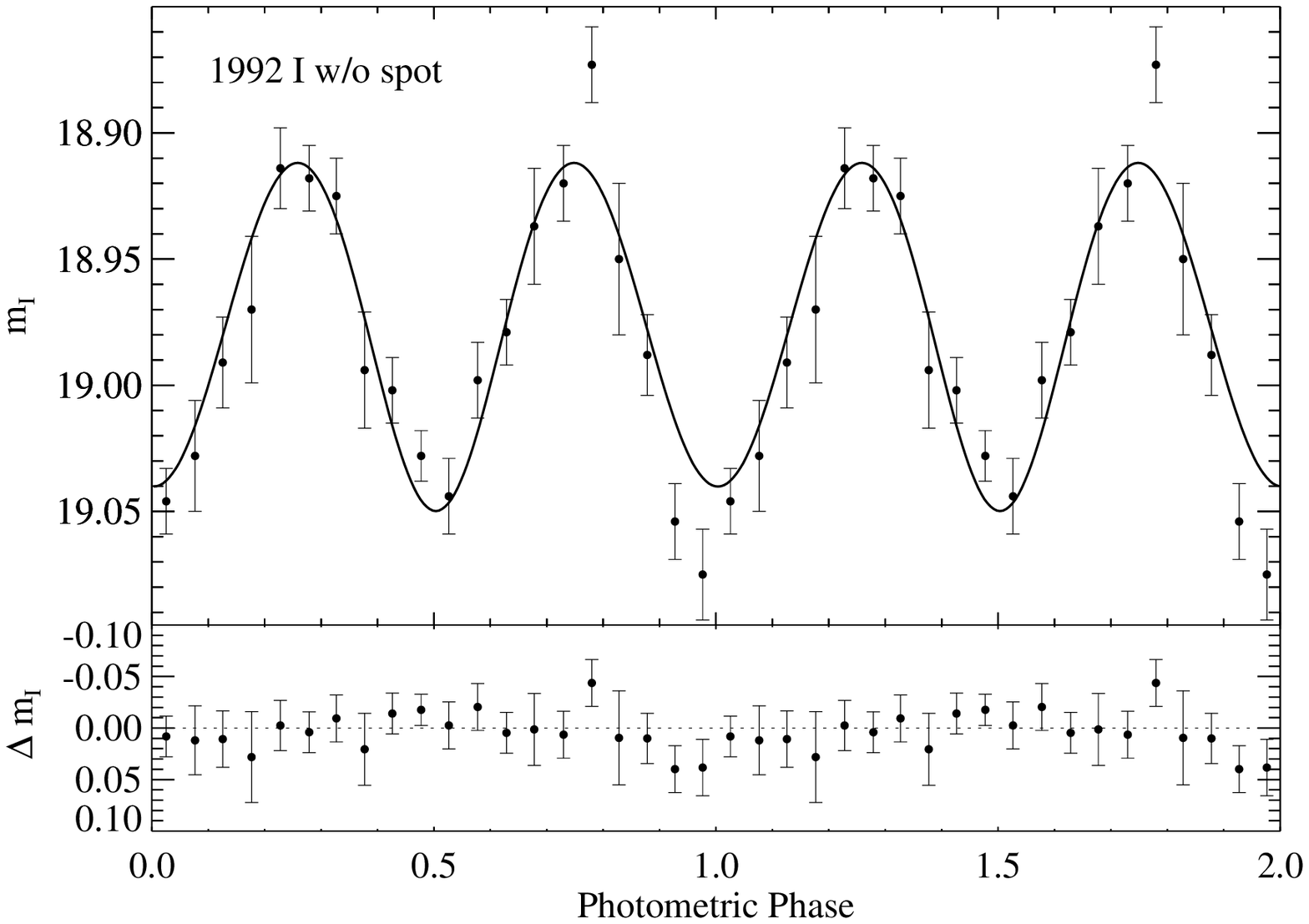}
    \caption{\footnotesize{The 1992 $I$-band light curve and the
    best-fit models with a disk spot (upper panel) and without a
    disk spot (lower panel). Two orbital cycles are plotted for clarity.} 
    \label{lc1_fig}}
\end{figure}%

\begin{figure}
    \centering
    \includegraphics[width=3.4in]{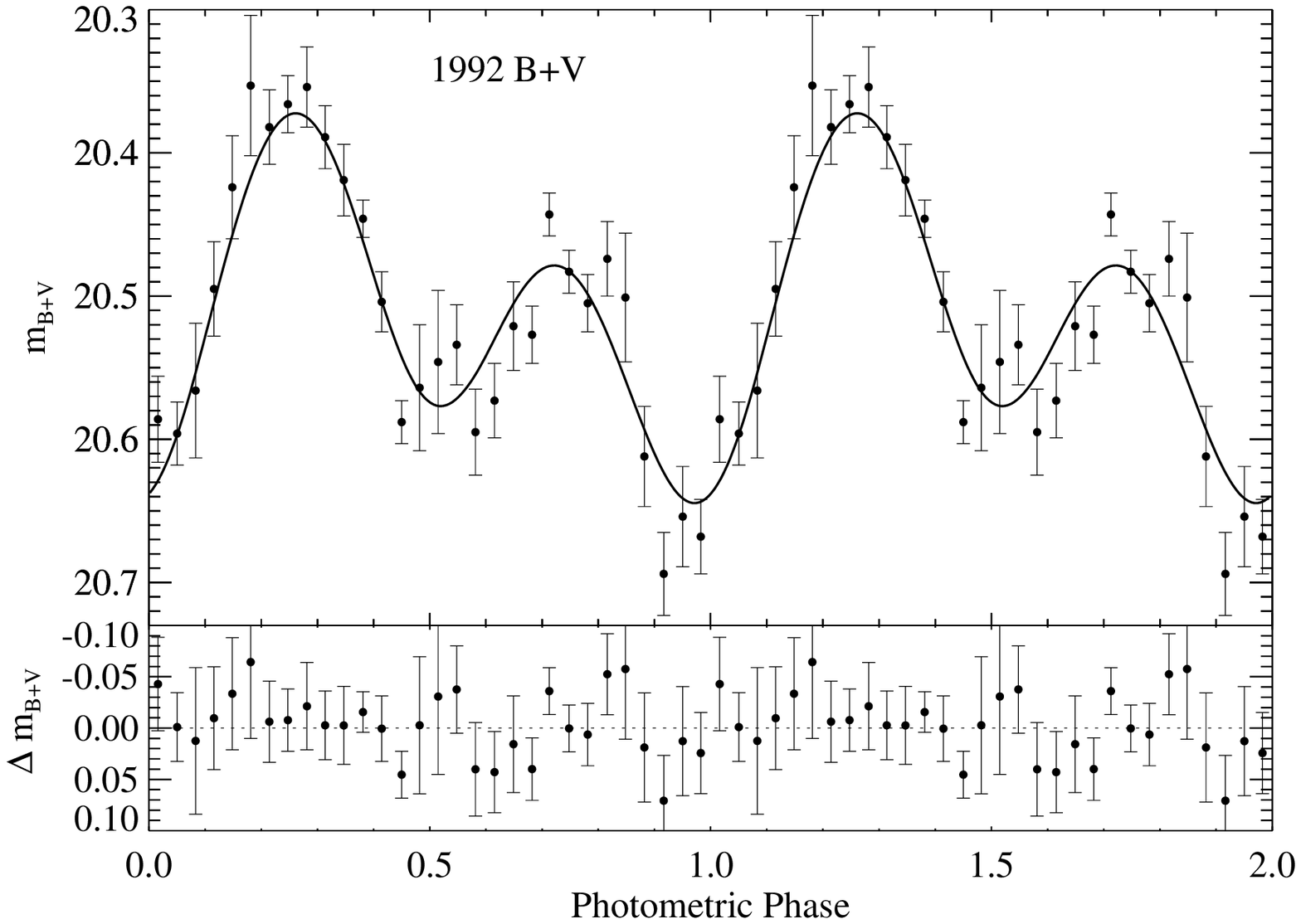}\\
    \includegraphics[width=3.4in]{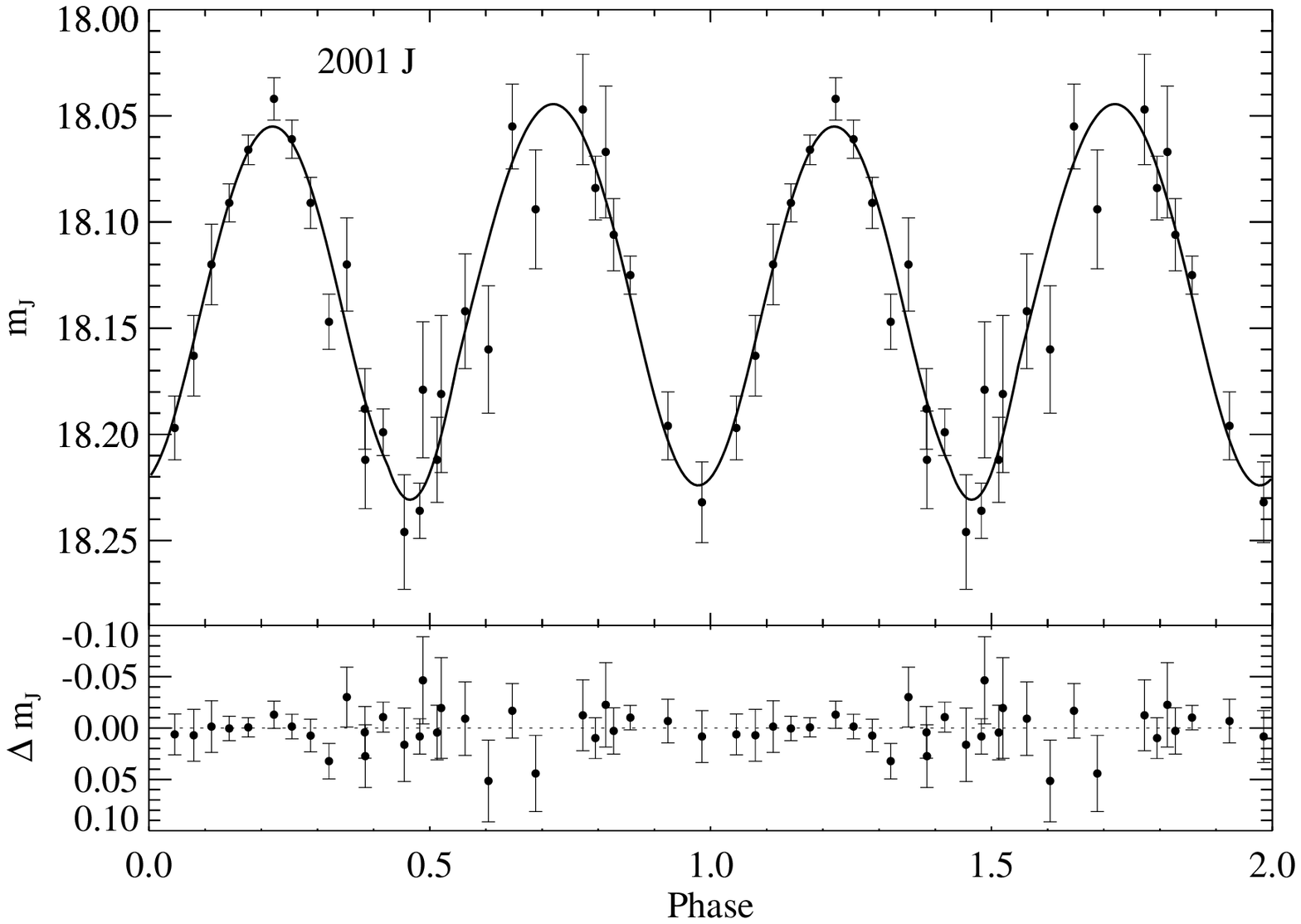}
    \caption{\footnotesize{The 1992 $B+V$-band light curve (upper
    panel) and G01 $J$-band light curve (lower panel), each with their
    own best-fit model that includes a disk spot. Two orbital cycles
    are plotted for clarity.} 
    \label{lc2_fig}}
\end{figure}%

The 1992 wide $I$-band light curve (see Fig.~\ref{lc1_fig}) of
\citet{orosz+1996} appears most desirable because it is least affected
by aperiodic flickering. Meanwhile, we also choose to model the
$B+V$-band light curve (see upper panel of Fig.~\ref{lc2_fig}), which
was obtained during the same year, even though it is distorted,
presumably by emission from a spot on the disk (see \S\ref{intro}).  We
further selected the $J$-band light curve of \citet{gelino+2001}, which
appears to be relatively free of aperiodic variations and dominated by
ellipsoidal modulation, while ignoring their relatively poor quality
$K$-band light curve.
\begin{deluxetable}{lcccc}
\tabletypesize{\footnotesize}
\tablecaption{Fraction of Disk Emission for Three Light Curves\label{veiling_table}}
\tablewidth{0pt}
\tablehead{ \colhead{Light Curve} &  \colhead{Mean $V$-band} & &
  \multicolumn{2}{c}{Disk Fraction (\%)} \\
\cline{4-5}
\colhead{} & \colhead{Magnitude} & & \colhead{$V$-band} & \colhead{$I$-band}
}
\startdata
2009 $V$ & $19.85\pm0.05$ & & $56.7\pm1.4$ & $42.1\pm2.8$ \\
1992 $B+V$, $I$ & $20.51\pm0.07$ & & $20.5^{+6.1}_{-6.6}$ &
$12.5^{+4.2}_{-4.3}$ \\
2001 $J$ & $20.83\pm0.45$ & & $<29.5$ & $<18.8$
\enddata
\end{deluxetable}

\begin{deluxetable*}{lccl}[t]
\tabletypesize{\footnotesize}
\tablecaption{List of ELC Model Parameters\label{parlist_table}}
\tablewidth{0pt}
\tablehead{ \colhead{Parameter} &  \colhead{Lower} & \colhead{Upper} &
  {Definition (unit)} \\ 
\colhead{} & \colhead{Bound} & \colhead{Bound} & \colhead{}
}
\startdata
$i$ & 30 & 80 & Inclination (deg) \\
$K_2$ & 395 & 420& Radial velocity semi-amplitude of the secondary (km s$^{-1}$) \\
$M_2$ & 0.3 & 1.5 & Mass of the secondary ($M_\odot$) \\
$\Delta\phi$ & -0.015 & 0.015 & Phase shift (deg) \\
$r_{\rm out}$ & 0.300 & 0.999 & Radius of the outer rim of the accretion disk\tablenotemark{a} \\
$\beta_{\rm rim}$ & 0.5 & 22.5 & Opening half-angle of the disk rim (deg)\\
$T_{\rm disk}$ & 3000 & 49000 & Temperature of the inner rim of the disk (K)\\
$\xi$ & -0.99 & -0.10 & Power-law index of the disk temperature profile \\
$s_{\rm spot}$ & 0.6 & 15 & The scale applied to obtain the disk spot
temperature \\
$r_{\rm cut}$ & 0.05 & 1.0 & Cut-off radius of the disk spot\tablenotemark{b} \\
$\theta_{\rm spot}$ & 0 & 360 & Azimuth of the disk spot (deg) \\
$w_{\rm spot}$ & 3 & 90 & Angular size of the disk spot (deg) \\
\hline
$M$ & \nodata & \nodata & Mass of the black hole ($M_\odot$) \\
$K_1$ & \nodata & \nodata & Radial velocity semi-amplitude of the
Black Hole (km s$^{-1}$)\\
$\log g_2$ & \nodata & \nodata & Surface gravity of the secondary (cm~s$^{-2}$)
\\
$R_2$ & \nodata & \nodata & Radius of the secondary ($R_\odot$)\\
$a$ & \nodata & \nodata & Separation between the centers of the black
hole and the secondary ($R_\odot$)\\
$v\sin i$ & \nodata & \nodata & Rotational Velocity of the secondary (km~s$^{-1}$)
\enddata
\tablecomments{The first set of parameters and their bounds are fit
parameters, and the second set are derived parameters.}
\tablenotetext{a}{In units of the effective Roche lobe radius of the black hole.}
\tablenotetext{b}{In units of the disk radius.}
\end{deluxetable*}

\subsection{Constraining the Fraction of Disk Emission}\label{lc:veil}

Although simultaneous spectroscopic data were not obtained for our
selected light curves, we are still able to constrain the fraction of
disk emission (referred to hereafter as the disk fraction) using the
results derived from our simultaneous spectroscopic/photometric campaign
in 2009 presented in \citetalias{wu+2015b}.  Following
\citet{cantrell+2010}, we make the reasonable assumption that the flux
from the secondary star remains constant, and that the change in
brightness of the system originates solely from the accretion disk.

During our 2009 observations, the mean $V$ magnitude of \nmus\ was
$19.85\pm0.05$, with the accretion disk contributing $56.7\pm1.4$\% of
the total light \citepalias{wu+2015b}.  This implies a magnitude for the
secondary star alone of $m_{\rm V}=20.76\pm0.06$.  During 1992, the mean $V$
magnitude of \nmus\ was $20.51\pm0.07$ \citep{orosz+1996}, just
$0.25\pm0.09$ mag brighter than the star alone.  Thus, in 1992
the disk fraction was only $20.5^{+6.1}_{-6.6}$\% (Table~\ref{veiling_table}).

With these $V$-band results in hand, we estimated the disk fraction for
the $I$-band using the spectral energy distribution model presented in
\S4.2 of \citetalias{wu+2015b}.  The model assumes a blackbody spectrum
with an effective temperature of 4400~K for the stellar component and a
power-law spectrum for the disk component (see Fig.~7 of
\citetalias{wu+2015b}). Extrapolating the power-law component, we obtain
for 1992 a disk fraction of $12.5^{+4.2}_{-4.3}$\% in the $I$-band.

Using the same procedure, we estimated the disk fraction for the 2001
$J$-band light curve (Table~\ref{veiling_table}).
However, in this case we only have a crude
estimate of the mean $V$-band magnitude of \nmus\ because
\citet{gelino+2001} report just a single measurement of $V$-band
magnitude, $20.83\pm0.06$, with no time-stamp.  Thus, the orbital phase
of the observation is unknown, and we therefore adopt an uncertainty of
${\Delta}m_{\rm V} = 0.45$ mag, i.e., the full amplitude of the 2009
$V$-band light curve.  Using this magnitude and our model, we constrain
the disk fractions in the $V$- and $I$-bands in 2001 to be $<29.5\%$ and
$<18.8\%$, respectively.  The $J$-band disk fraction is computed
separately using a stellar-atmosphere model during the modeling of the
light curve, which is described in the next section.

\begin{deluxetable*}{lcccccccc}
\tabletypesize{\footnotesize}
\tablecaption{Best-fit Parameters of ELC Models\label{parfit_table}}
\tablewidth{0pt}
\tablehead{ \colhead{Parameter} & & \multicolumn{3}{c}{1992~$I$} & &
  \colhead{1992~$B+V$ and $I$} & & \colhead{2001~$J$}\\ 
\cline{3-5}
\colhead{} & & \colhead{w/ spot} & & \colhead{w/o spot} & & \colhead{} &
& \colhead{}
}
\startdata
$i$~(deg) & & $43.2^{+2.1}_{-2.7}$ & & $40.5^{+2.3}_{-2.5}$ & & $40.7^{+3.4}_{-2.8}$& &$40.8^{+6.1}_{-2.6}$ \\
$K_2$~(km s$^{-1}$)& &$407.0^{+2.1}_{-2.3}$ & & $406.9^{+2.3}_{-2.2}$ & & $406.6^{+2.2}_{-1.8}$ & & $406.5^{+2.5}_{-1.7}$\\
$M_2$~($M_\odot$) & & $0.89^{+0.18}_{-0.11}$ & & $1.05^{+0.21}_{-0.15}$& & $1.03^{+0.26}_{-0.16}$ & & $1.02^{+0.21}_{-0.28}$\\
$\Delta\phi$ & & $0.006^{+0.008}_{-0.007}$ & & $0.003^{+0.010}_{-0.009}$ & & $>0.007$ & & $<0.008$\\
$r_{\rm out}$ & & $0.58$\tablenotemark{a} & & \nodata\tablenotemark{b} & & $>0.40$ & & $>0.76$\\
$\beta_{\rm rim}$~(deg)& & $6.4^{+5.8}_{-1.9}$ & & $10.7$ & & $>13.4$& & $>13.2$\\
$T_{\rm disk}$~($10^4$~K) & & $<1.77$ & &$>0.37$ & & $<1.49$& & $>0.80$\\
$\xi$ & & $-0.69^{+0.37}_{-0.12}$& &$>-0.68$ & & $-0.45^{+0.17}_{-0.22}$& & $-0.76^{+0.27}_{-0.11}$\\
$s_{\rm spot}$ & & $>3.3$ & & \nodata\tablenotemark{c} & & $>3.9$ & & $>3.4$ \\
$r_{\rm cut}$ & & $<0.26$ & & \nodata\tablenotemark{c} & & $<0.37$ & & $0.35^{+0.06}_{-0.26}$\\
$\theta_{\rm spot}$~(deg) & & $10^{+29}_{-26}$ & & \nodata\tablenotemark{c} & &$304^{+7}_{-15}$ & & $74^{+6}_{-15}$\\
$w_{\rm spot}$~(deg) & & $17^{+20}_{-12}$ & & \nodata\tablenotemark{c} & &$<26$ & & $33^{+14}_{-15}$\\
\hline\\
$M$~($M_\odot$) &  & $11.0^{+2.1}_{-1.4}$ & &$12.9^{+2.2}_{-1.7}$ & &$12.7^{+3.1}_{-1.8}$ & & $12.6^{+2.4}_{-3.5}$\\
$K_1$~(km s$^{-1}$) &  & $32.8^{+2.5}_{-1.9}$ & & $33.0^{+2.5}_{-2.4}$& & $33.0^{+2.4}_{-2.3}$& & $32.8\pm2.3$\\
$\log g_2$~(cm s$^{-1}$) &  & $4.34^{+0.03}_{-0.02}$ & &$4.36^{+0.03}_{-0.02}$ & &$4.36\pm0.03$ & & $4.36^{+0.03}_{-0.05}$\\
$R_2$~($R_\odot$) &  & $1.06^{+0.07}_{-0.04}$& &$1.12^{+0.07}_{-0.05}$ & &$1.11^{+0.09}_{-0.06}$ & & $1.11^{+0.07}_{-0.11}$ \\
$a$~($R_\odot$) &  & $5.49^{+0.32}_{-0.24}$ & &$5.79^{+0.32}_{-0.26}$& & $5.76^{+0.44}_{-0.29}$ & &$5.75^{+0.35}_{-0.60}$\\
$v\sin i$~(km s$^{-1}$) &  & $84.8^{+2.3}_{-1.9}$ & &
$84.9^{+2.3}_{-2.2}$& & $84.9^{+2.3}_{-2.2}$& & $84.7\pm2.2$\\
\hline\\
$\chi^2_\nu\ (\nu)$ & & $1.03\ (8)$ & & $1.10\ (12)$ & & $0.88\ (38)$ & &
$0.84\ (17)$
\enddata
\tablecomments{The quoted uncertainties or upper/lower limits are at the
  $1\sigma$ level of confidence.}  
\tablenotetext{a}{The quoted value
  corresponds to the model with minimum $\chi^2$. However, no meaningful
  uncertainty range can be given because the $\chi^2$ curve is flat.}
\tablenotetext{b}{Parameter is unconstrained in the absence of a spot.}
\tablenotetext{c}{These spot parameters are irrelevant for this model.}
\end{deluxetable*}

\section{Dynamical Modeling}\label{model}

\subsection{Description of the Eclipsing Light Curve Model}\label{model:elc}

In constraining the inclination $i$ of \nmus, we model our three selected
light curves using the Eclipsing Light Curve \citep[ELC;][]{orosz+2000}
code.  The ELC code generates model light curves that include
contributions from both the secondary star and the accretion disk,
including an allowance for emission from a hot spot on the disk.  The
secondary star is assumed to be in a circular orbit, rotating
synchronously, and filling its Roche lobe.  We fix the effective
temperature of the secondary to be $T_2=4400\pm100$~K based on the
results of our cross correlation analysis (see \S3.1 and Fig.~2 of
\citetalias{wu+2015b}); meanwhile our ELC models are insensitive to the
adopted value of $T_2$.  We ignore \xray\ heating because the quiescent
\xray\ luminosity of \nmus\ is minuscule ($L_{\rm X}\sim
4\times10^{31}$~erg~s$^{-1}$; \citealt{sutaria+2002}).
 
The fit and derived parameters of the ELC models are listed in
Table~\ref{parlist_table}.  There are four key fit parameters: the
inclination $i$, $K$-velocity of the secondary $K_2$, mass of the
secondary $M_2$ and a relative phase shift $\Delta\phi$.  Our measured
values of $K_2$ ($=406.8\pm2.7$~km~s$^{-1}$) and $v\sin i$
($=85.0\pm2.6$~km~s$^{-1}$) reported in \citetalias{wu+2015b} are used to
generate the prior probability distribution for each of the four
parameters.  The posterior probability distribution of the parameters
generated by the ELC code are required to agree closely with the
measured values, which serves as a consistency check on the modeling
process. We fix the orbital period to be $P=0.43260249(9)$~d (Table~2 of
\citetalias{wu+2015b}).

There are four additional fit parameters that characterize the emission
from the accretion disk: 1) the radius of the outer rim of the disk
$r_{\rm out}$, expressed in units of the effective Roche lobe radius of
the primary \citep{eggleton+1983}; 2) the half opening angle of the
accretion disk $\beta_{\rm rim}$; 3) the temperature of the inner edge
of the disk $T_{\rm disk}$; and 4) the power-law index of the
temperature profile along the disk radius $\xi$.  For those models that
include a hotspot along the outer rim of the accretion disk, there are
four additional fit parameters: $s_{\rm spot}$ (the scaling factor
used to derive the spot temperature from the disk temperature); $r_{\rm
  cut}$ (the inner cut-off radius of the spot); $\theta_{\rm spot}$ (the
azimuth of the disk spot relative to the line connecting the center of
the primary and secondary); and $w_{\rm spot}$ (the angular width of the
disk spot).

For each fit parameter, we set a range of physically and/or
geometrically reasonable values (see Table~\ref{parlist_table}), some of
which are based on the results in \citetalias{wu+2015b}.  For each set of
trial values of the parameters, and for a particular filter band, the
ELC code generates a model light curve.  The observed light curves are
fitted to the model curves using a variety of optimizing techniques
\citep[see details in \S5.2.1 of][]{orosz+2014}.  After a large number
of trials, the model giving the global minimum $\chi^2$ is
adopted. Other systemic parameters, such as the black hole mass $M$ and
the radius of the secondary $R_2$, are derived using this best-fit model
(see Table~\ref{parlist_table}).

\begin{figure*}[t]
    \centering
    \includegraphics[width=6in]{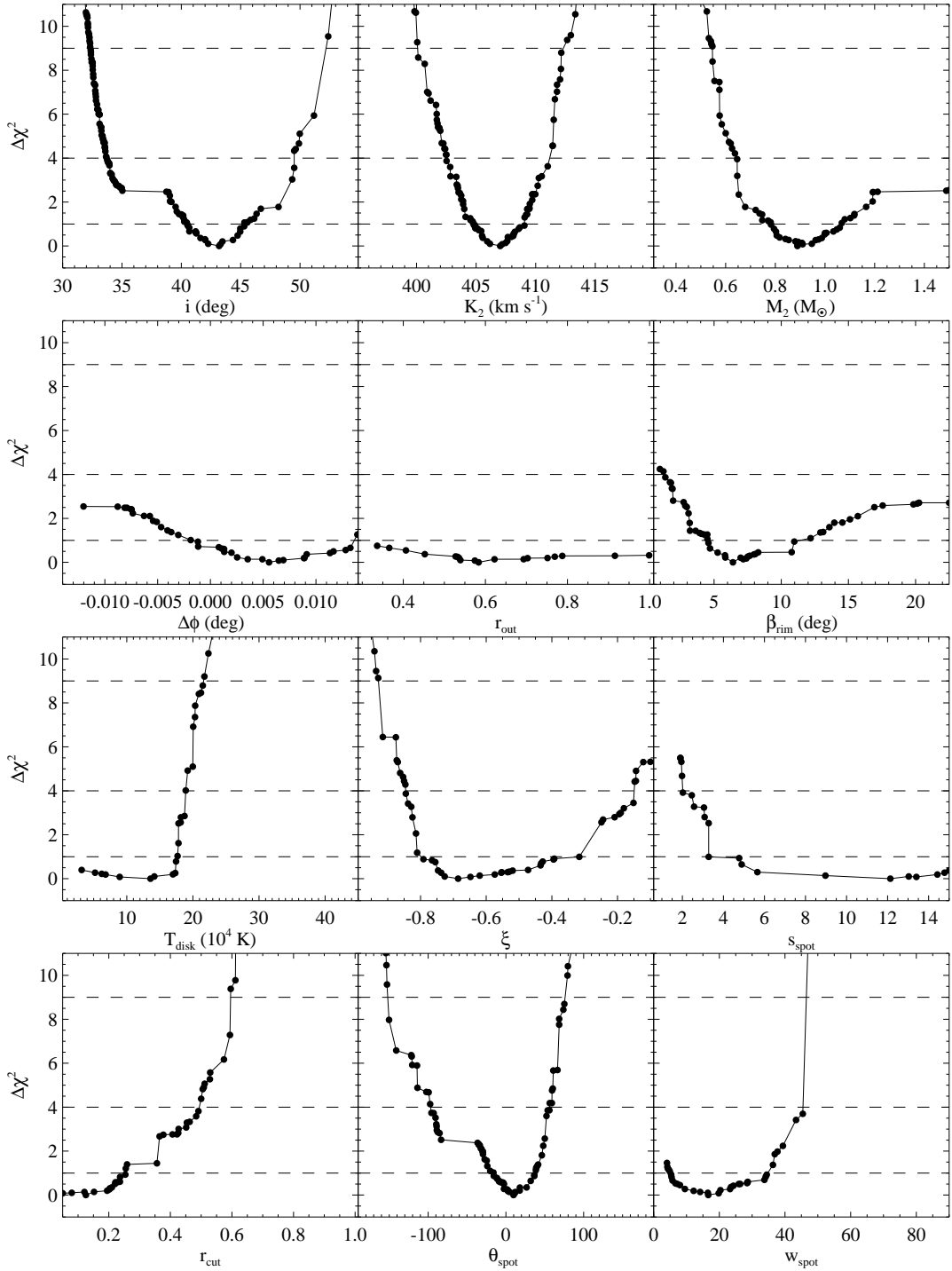}
    \caption{\footnotesize{The $\chi^2$ curves for the fit
    parameters. The dashed
    horizontal lines illustrate the $1\sigma$, $2\sigma$ and $3\sigma$
    uncertainty ranges.}
    \label{chi21_fig}}
\end{figure*}%

\begin{figure*}[t]
    \centering
    \includegraphics[width=6in]{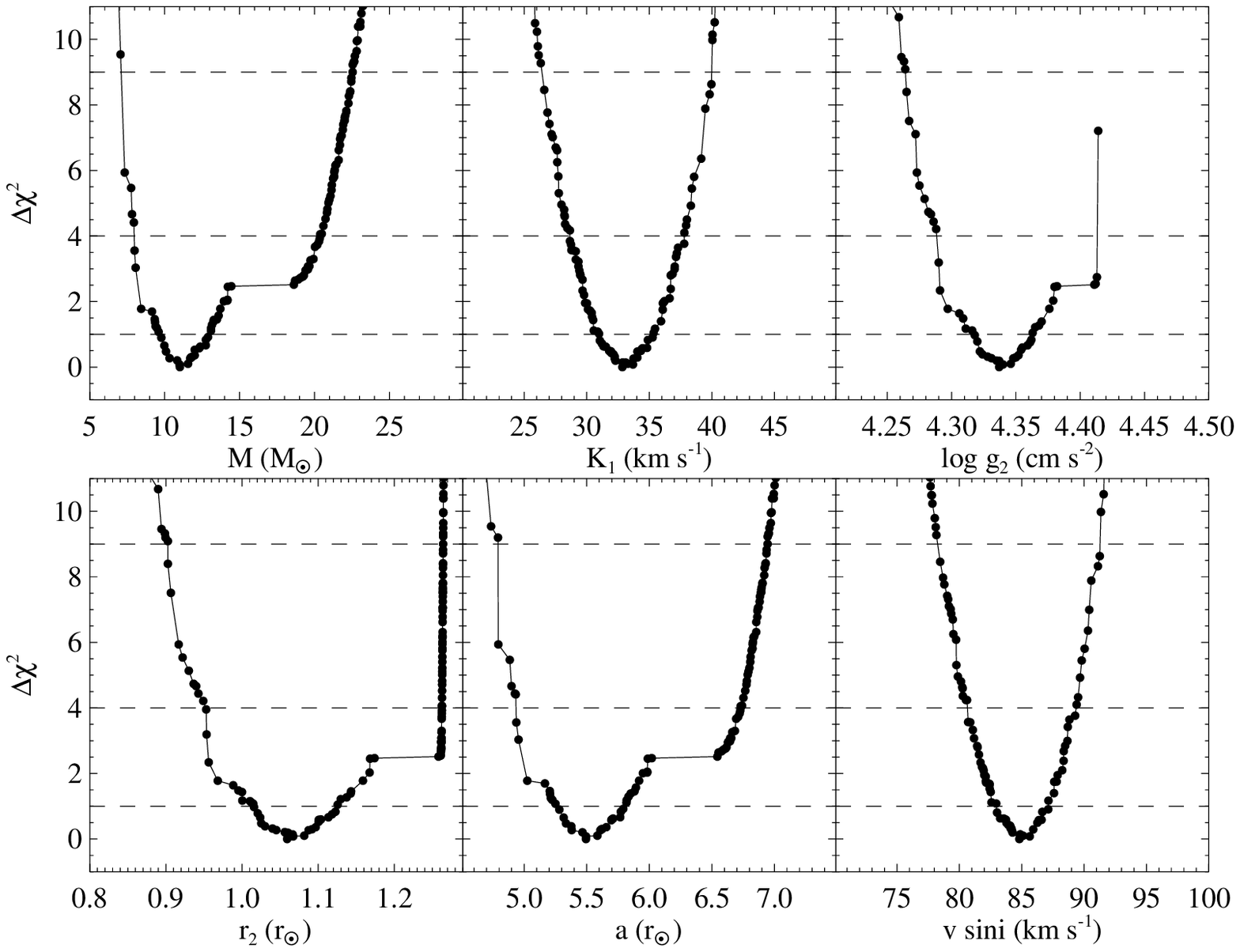}
    \caption{\footnotesize{Same as Fig.~\ref{chi21_fig}, but for the derived
    parameters.}
    \label{chi22_fig}}
\end{figure*}%

\subsection{Light Curve Fitting}\label{model:fit}

As already indicated in \S\ref{lc:pick}, among the three selected light
curves the 1992 $I$-band light curve appears to be the most favorable:
it is weakly affected by aperiodic flickering, its disk fraction is
minimal ($\approx12.5$\%), and it is much less distorted than its
companion $B+V$-band curve.  Compared to the 2001 $J$-band light curve,
the constraint on the disk fraction for the $I$-band light curve is
tighter and the disk was significantly fainter in 1992 ($\sim0.3$~mag;
\S\ref{lc:smarts}).  We therefore choose the 1992 $I$-band light curve
as our primary data set. 

We generated a total of 2.5 million model light curves using the ELC
code for the 1992 $I$-band data. The models include emission from the
secondary star and from the disk, which is comprised of a steady
component and a periodically variable component due to a hotspot.  The
model light curve that yields the minimum $\chi^2$ is identified.  The
corresponding fit parameter values are listed in the second column of
Table~\ref{parfit_table}.  For all 2.5 million models, we plotted the
parameter value vs. $\Delta\chi^2$.  In
Figs.~\ref{chi21_fig}~and~~\ref{chi22_fig}, we show respectively for
the fit parameters and the derived parameters only the outer envelope
defined by the mass of points.  The three dashed lines shown in each
panel correspond to $\Delta\chi^2=1,4,9$. The points of intersection of
these lines with the envelope define respectively the $1\sigma$,
$2\sigma$, $3\sigma$ uncertainty range of the parameter in question.
For some fit parameters (e.g., $T_{\rm disk}$) there is only a one-sided
constraint; in these cases, in Table~\ref{parfit_table} we quote
$1\sigma$ lower/upper limits.

\begin{figure}[t]
    \centering
    \includegraphics[width=3.5in]{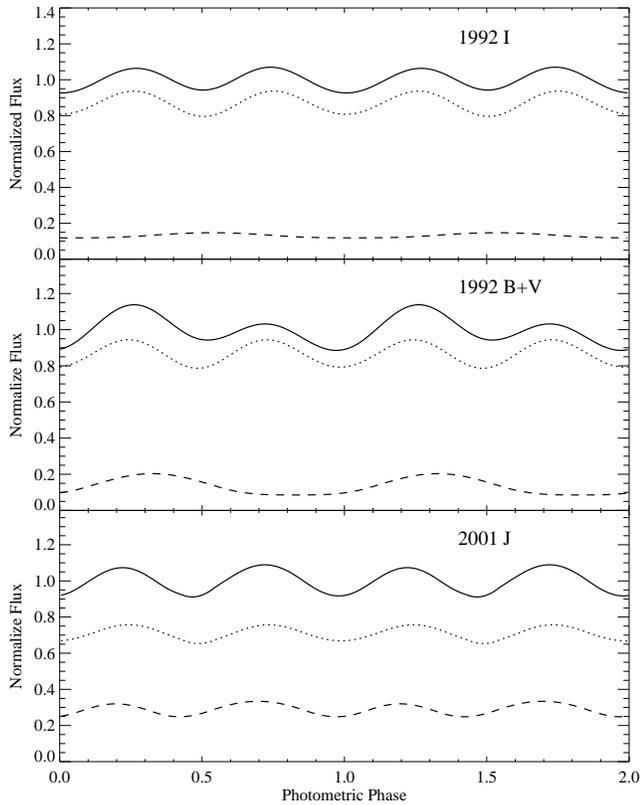}
    \caption{\footnotesize{The model light curve for each filter band
        (solid line) shown decomposed into the stellar component (dotted
        line) and the disk component (dashed line). The flux is in
        linear units and the average level for the total model is
        normalized to unity. Two orbital cycles are plotted for
        clarity.}
    \label{lcde_fig}}
\end{figure}

For our primary data set, the 1992 $I$-band light curve, the observed
and model light curves, and fit residuals, are shown in the upper panel
of Fig.~\ref{lc1_fig}.  The model curve is itself shown decomposed into
contributions from the secondary and from the disk in the top panel of
Fig.~\ref{lcde_fig}.  As this figure makes clear, both the constant and
variable components of disk emission are relatively unimportant.

We also fitted the 1992 $I$-band light curve with models with no hotspot
component.  The values of the fit parameters, which are given in the
third column of Table~\ref{parfit_table}, are in good agreement with
those obtained for the model that includes a spot. However, the quality
of the fit is significantly poorer, as shown by an inspection of the fit
residuals, which show evidence of modulation at the orbital period
(lower panel of Fig.~\ref{lc1_fig}).

We further fitted the 1992 $B+V$-band and $I$-band light curves jointly,
and the 2001 $J$-band light curve, including the hotspot in both cases,
and we find results that are consistent with those obtained for our
primary data set.  The results are listed in the fourth and fifth
columns of Table~\ref{parfit_table}, respectively.  The disk fractions
we used (including both the constant and hotspot emission) are those
given in Table~\ref{veiling_table}.  In both cases, the best-fit
parameters are consistent with those obtained for our primary data set,
the 1992 $I$-band light curve.  Furthermore, in all cases the values of
$K_2$ and $v\sin i$ returned by the models agree well with our measured
values given in \citetalias{wu+2015b}.  Thus, the fit results for all
three light curves are consistent, and the models in all cases are in
good agreement with our dynamical data.

An inspection of Fig.~\ref{lcde_fig}, which shows the model light curves
decomposed into stellar and disk components, makes clear the principal
virtue of the 1992 $I$-band light curve (top panel), namely, that its
ellipsoidal component is minimally contaminated by the hotspot
component, as well as being minimally diluted by the steady disk
component. Confirmation that disk contamination was minimal during this
period is provided by comparing the results of two Doppler tomographic
studies: No hotspot was detected in a 1994--1995 tomogram
\citep{casares+1997}, while a prominent hotspot, located where the gas
stream strikes the disk, is present in a 2009--2010 tomogram
\citep{peris+2015}.

\begin{figure}[t]
    \centering
    \includegraphics[width=3.5in]{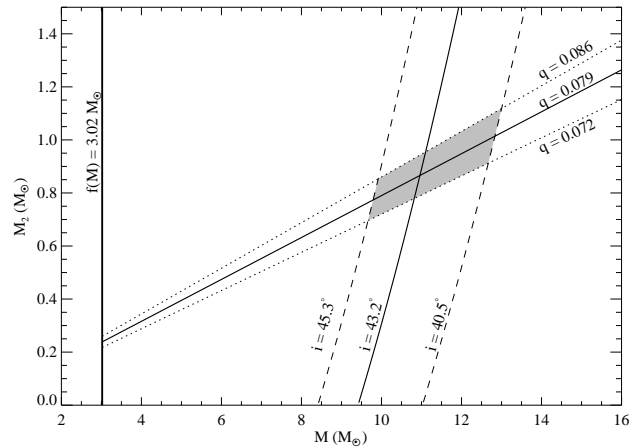}
    \caption{\footnotesize{Constraints on the masses of the primary and
        secondary shown in the $M-M_2$ plane; the $1\sigma$ range of
        uncertainty is indicated by the gray-shaded region.} 
    \label{massdia_fig}}
\end{figure}

Table~\ref{orbital_table} summarizes our final, key results for \nmus.
The mass of the black hole is $M=11.0^{+2.1}_{-1.4}\ M_\odot$.  The mass
of the secondary star is slightly less than a solar mass
($M_2=0.89^{+0.18}_{-0.11}\ M_\odot$).  The relation between the two
masses, given our constraints on $q$ and $i$, are summarized in
Fig.~\ref{massdia_fig}.  The vertical solid line on the left marks the
hard lower limit on $M$ imposed by our measurement of the mass function,
and the slant solid lines show the constraints imposed by our measured
values of $q$ and $i$, while the dotted and dashed lines indicate
$1\sigma$ errors, respectively.  The gray-shaded area defines the
$1\sigma$ range of uncertainty in this $M$--$M_2$ diagram.  In the
following section we will use the radius of the secondary
($R_2=1.06^{+0.07}_{-0.04}~R_\odot$), which is essentially the solar
value, to estimate the distance of \nmus.

\bibpunct[; ]{(}{)}{;}{a}{}{,}

The black hole mass we find is significantly greater than the
$M=6.95\pm0.6\ M_\odot$ value reported by \citet[also see
  \citealt{gelino+2004}]{gelino+2001}, which is 
primarily because our estimate of inclination is lower, $43.2^\circ$
vs. $54^\circ$.  That Gelino et al.\ obtained a higher inclination while
ignoring the disk emission is contrary to the usual expectation because
ignoring a pedestal of light causes one to underestimate the inclination
\citep[e.g.,][]{kreidberg+2012}.  The present case is unusual because
the pedestal of light is relatively unimportant and the effect of the
hotspot is dominant.  Our model shows that about half the total
modulation of the 2001 $J$-band light curve is due to the hot spot
component (Fig.~\ref{lcde_fig}).  Gelino et al.\ did not include this
component and they therefore overestimated the inclination.

\begin{center}
\begin{deluxetable}{lcl}
\tabletypesize{\footnotesize}
\tablecaption{Key Parameters for \nmus\ \label{orbital_table}}
\tablewidth{0pt}
\tablehead{ \colhead{Parameter} & \colhead{} & \colhead{Value}
}
\startdata
Mass function $f(M/M_\odot)$ & & $3.02\pm0.06$ \\
Mass ratio $q$ & & $0.079\pm0.007$ \\
Inclination $i$ (deg) & & $43.2^{+2.1}_{-2.7}$ \\
Black hole mass $M$ ($M_\odot$) & & $11.0^{+2.1}_{-1.4}$ \\
Secondary mass $M_2$ ($M_\odot$) & & $0.89^{+0.18}_{-0.11}$ \\
Secondary radius $R_2$ ($R_\odot$) & & $1.06^{+0.07}_{-0.04}$ \\
Separation $a$ ($R_\odot$) & & $5.49^{+0.32}_{-0.24}$ \\
Distance $D$ (kpc) & & $4.95^{+0.69}_{-0.65}$
\enddata
\tablecomments{The quoted uncertainties are at the $1\sigma$ level of
  confidence.}
\end{deluxetable}
\end{center}




\section{Distance of \nmus}\label{dist}

We follow the methodology outlined in \S2 of \citet{barret+1996} to
estimate the distance of \nmus. We first determined the hypothetical
absolute magnitude of the secondary star $\mathcal{M_{\rm V}}$ as viewed at its 
surface (i.e., at a distance $D = R_2$, rather than the canonical
distance of 10 pc) using the most recent stellar atmosphere models of
late-type stars.  Then, using our apparent $V$-band magnitude corrected
for reddening, and our estimate of the star's radius, we computed the
distance of \nmus.

We utilize the BT-Settl stellar atmosphere models developed by
\citet{allard+2012a,allard+2012b}\footnote{The most recent model library
  is available online at
  \url{https://phoenix.ens-lyon.fr/Grids/BT-Settl/CIFIST2011\_2015/.}}.
These models provide the magnitude $\mathcal{M_{\rm V}}$ of the star at
the stellar surface for a variety of filter bands over a wide grid of
effective temperatures (in steps of 100~K) and surface gravities (in
logarithmic steps of 0.5).  For \nmus, the effective temperature is
$T_2=4400\pm100$~K (\S\ref{model:elc}) and the surface gravity is $\log
g_2=4.34^{+0.03}_{-0.02}$ (Table~\ref{parfit_table}).  Averaging the
entries in the table for models in the range $\log g=4.0-4.5$ with
$T_2$ fixed at 4400~K, we find $\mathcal{M_{\rm V}}=-36.72$.

The uncertainty in $\mathcal{M_{\rm V}}$ associated with this range of
$\log g$ and the 100~K uncertainty in $T_2$ is 0.19~mag.  It is
also important to assess the uncertainty in the model of
\citeauthor{allard+2012a} For this purpose, we selected several stars
with values of $T_2$ and $\log g$ close to those of the secondary for
which $\mathcal{M_{\rm V}}$ can be deduced purely from observational
data \citep{boyajian+2012}.  The data for these stars generally agree
with the models of \citeauthor{allard+2012a} to within 0.1~mag.
We combine this uncertainty in quadrature with 0.19~mag to obtain
the value we adopt for the absolute surface magnitude of the secondary:
$\mathcal{M_{\rm V}}=-36.72\pm0.21$.

The apparent $V$-band magnitude of \nmus\ is $19.85\pm0.05$ in 2009;
correcting for the disk contribution $(56.7\pm1.4)\%$, the magnitude of
the secondary alone is $m_{\rm V}=20.76\pm0.06$.  Next, we correct for
reddening, relying on two observations of the 2200~\AA\ dust feature in
the ultraviolet spectrum of \nmus.  \citet{cheng+1992} reported
$E(B-V)=0.287\pm0.004$ based on observations made using the Faint Object
Spectrograph (FOS) onboard the {\it Hubble Space Telescope}, while
\citet{shrader+1993} obtained a consistent result, $E(B-V)=0.30\pm0.05$
based on observations made using the {\it International Ultraviolet
  Explorer}.  We take $E(B-V)=0.29\pm0.05$, adopting the larger error 
bar to cover the uncertainty in
converting $E(B-V)$ to $A_{\rm V}$ using the standard extinction law:
$A_V=R(V)\times E(B-V)$ with $R(V)=3.1$ \citep{cardelli+1989}.  We
therefore conclude that the reddening of \nmus\ is $A_{\rm
  V}=0.90\pm0.16$ and the dereddened $V$-band magnitude of the secondary
is $m_{\rm V_0}=19.86\pm0.17$.

The distance $D$ can then be obtained from the equation 
\begin{eqnarray}
5\log(D/R_2)=m_{\rm V_0}-\mathcal{M_{\rm V}}=56.58\pm0.27,
\end{eqnarray}
where $R_2=1.06^{+0.07}_{-0.04}\ R_\odot$ is the
radius of the secondary (\S\ref{model:fit}).  We therefore conclude that
the distance of \nmus\ is $D=4.95^{+0.69}_{-0.65}$~kpc.  This value is
consistent with most literature estimates, which are generally based on
the methodology we employ; e.g., $5.1$~kpc \citep{gelino+2001},
$5.0\pm1.0$~kpc \citep{esin+1997}, $5.5\pm1.0$~kpc \citep{orosz+1996},
and $>4$~kpc \citep{king+1996}.  Using NIR photometric data, while
ignoring the disk contribution, \citet{shahbaz+1997} obtained a lower
estimate of $D=2.8$--4.0~kpc.


\section{Summary \& Discussion}\label{discuss}

Combining the various results presented herein with those in
\citetalias{wu+2015b}, we have obtained three principal results:

\begin{enumerate}

\item We present the first evidence that BHSXTs brighten gradually and
  steadily between outburst cycles.  For \nmus, SMARTS data show the
  rate to be $0.0357\pm0.0006$~mag~yr$^{-1}$ in the $V$-band and
  $0.0199\pm0.0004$~mag~yr$^{-1}$ in the $I$-band for the period 2003 to
  2015 (sans 2010). Between the time \nmus\ returned to quiescence
  following its 1991 outburst and today, the total brightening in the
  $V$-band is $\approx0.8$~mag.  This result provides support for the
  disk instability model and the work of
  \citet{cantrell+2008,cantrell+2010} on optical states and state
  changes.  It also underscores the importance of obtaining light
  curve data early in the quiescent phase for use in constraining the
  systemic inclination via ellipsoidal modeling. 
\item By modeling three archival optical/NIR light curves of the highest
  quality using the ELC code, we determined the systemic inclination
  of \nmus\ to be $i=43.2^{+2.1}_{-2.7}$ deg.  Our measured value of the
  black hole mass is $M=11.0^{+2.1}_{-1.4}\ M_\odot$, while the mass of the
  secondary is $M_2=0.89^{+0.18}_{-0.11}\ M_\odot$.
\item Based on our determination of the radius of the secondary,
  $R_2=1.06^{+0.07}_{-0.04}~R_\odot$, we estimate the distance of \nmus\
  to be $D=4.95^{+0.69}_{-0.65}$~kpc.  We will use this estimate of $D$,
  along with our estimates of $i$ and $M$, archival X-ray data, and the
  continuum-fitting method to estimate the spin of the black hole, a
  result that will be presented in a companion paper (Chen et al. 2015).

\end{enumerate}

Among transient X-ray sources, the black hole mass in \nmus\ is high,
deviating by $\sim3\sigma$ from the narrowly-distributed mass
distribution of BHSXTs centered at $7.8\pm1.2\ M_\odot$
\citep{ozel+2010,farr+2011,mcclintock+2014}.  The two rivals are the
black holes in GRS~1915$+$105 \citep[$M=12.4^{+2.0}_{-1.8}\
M_\odot$;][]{reid+2014} and in GS~2023+338 \citep[$M=12 \pm 2
M_\odot$;][but see \citealt{khargharia+2010}]{charles+2006}.  The general lack of
massive black holes among BHSXTs can be explained by the evolutionary
paths of low-mass \xray\ binaries during which the star loses
substantial mass before the supernova explosion \citep{ozel+2010}.  The
black hole in \nmus\ may be spinning relatively rapidly as the result of
accretion torques, if the black hole's birth mass is significantly less
than the current mass \citep{fragos+2015}.

We note three lessons learned for future attempts to obtain robust
measurements of the masses of black holes in BHSXTs based on our results
and on the results of earlier studies
\citep[e.g.,][]{cantrell+2008,cantrell+2010,kreidberg+2012}: 1) perform
the spectroscopic and photometric monitoring observations simultaneously
in order to accurately correct for disk emission; 2) use only
passive-state light curve data for which the non-stellar emission is
free of strong flickering and can be well-modeled by a steady disk
component plus a hotspot component; and 3) obtain optical monitoring
data soon after the system returns to quiescence, when the disk emission
is minimal.

Concerning the first of the three points -- the need for simultaneity --
this is particularly important for short-period systems because the disk
component is often dominant. Furthermore, as an inspection of
Fig.~\ref{smarts_fig} makes clear, this component not only varies
from season to season, but it varies on much shorter
timescales as well.  The key methodology we employed was insuring that
our spectroscopic and photometric data were obtained strictly
simultaneously.  Although the $V$-band light curve we obtained in 2009
was unsuitable for ellipsoidal modeling because the system was in the
active state, we were nevertheless able to use it (and the spectroscopic
data) to very precisely constrain the disk fraction for our selected
archival light curves (\S\ref{lc:veil}).

In closing, we comment briefly on ellipsoidal studies of short-period
BHSXTs by other groups (see also
\citealt{kreidberg+2012}). \citet{cantrell+2010} obtained robust
estimates of $i$ and $M$ for A0620$-$00 using nearly simultaneous
spectroscopic and photometric data to constrain the disk fraction. 
 \citet{khargharia+2013} likewise obtained
high-quality constraints on $i$ and $M$ for XTE~J1118$+$480 based on NIR
photometry and spectroscopy performed on the same night.  GS~2000$+$25
appears to be a very favorable system; the disk fraction is minimal, and
the orbital modulation almost purely ellipsoidal. However,
\citet{ioannou+2004} report relatively weak constraints on $i$ and $M$
because the disk fraction and the mass ratio $q$ were poorly
constrained.  For XTE~J1859$+$226, \citet{corral+2011} obtained
passive-state light curves, but their disk fraction is poorly
constrained because two years elapsed between their photometric and
spectroscopic observations.  The light curves of GRO~J0422$+$32 are
always observed to be dominated by aperiodic flickering;
\citet{reynolds+2007} conclude that attempts to constrain the mass of
the black hole are ``prone to considerable uncertainty'' because of the
contaminating effects of disk emission.  In conclusion, one must be
cautious of masses published for short-period BHSXTs based on
non-simultaneous photometry and spectroscopy, and doubly cautious of
studies that ignore the disk emission altogether
\citep[e.g.][]{gelino+2003}.


\begin{acknowledgments}
We thank D.~Steeghs, P.~Longa-Pe\~{n}a, P.~J.~Callanan, L.~C.~Ho,
P.~G.~Jonker, M.~T.~Reynolds, and M.~A.~P.~Torres for their
contributions to this work, which are reported in 
\citetalias{wu+2015b}. L.J.G acknowledges the support of the Chinese
Academy of Sciences  
through grant No. XDB09000000 (Emergence of Cosmological Structures)
from the Strategic Priority Research Program; of the National Natural
Science Foundation of China (grant No. 11333005); and of the National
Astronomical Observatories of China (grant No. Y234031001). 
\end{acknowledgments}



\bibliographystyle{apj}
\bibliography{master}

\begin{thebibliography}{}
\expandafter\ifx\csname natexlab\endcsname\relax\def\natexlab#1{#1}\fi

\bibitem[{{Allard} {et~al.}(2012{\natexlab{a}}){Allard}, {Homeier}, \&
  {Freytag}}]{allard+2012a}
{Allard}, F., {Homeier}, D., \& {Freytag}, B. 2012{\natexlab{a}}, Royal Society
  of London Philosophical Transactions Series A, 370, 2765

\bibitem[{{Allard} {et~al.}(2012{\natexlab{b}}){Allard}, {Homeier}, {Freytag},
  \& {Sharp}}]{allard+2012b}
{Allard}, F., {Homeier}, D., {Freytag}, B., \& {Sharp}, C.~M.
  2012{\natexlab{b}}, in EAS Publications Series, Vol.~57, EAS Publications
  Series, ed. C.~{Reyl{\'e}}, C.~{Charbonnel}, \& M.~{Schultheis}, 3--43

\bibitem[{{Barret} {et~al.}(1996){Barret}, {McClintock}, \&
  {Grindlay}}]{barret+1996}
{Barret}, D., {McClintock}, J.~E., \& {Grindlay}, J.~E. 1996, \apj, 473, 963

\bibitem[{{Beer} \& {Podsiadlowski}(2002)}]{beer+2002}
{Beer}, M.~E., \& {Podsiadlowski}, P. 2002, \mnras, 331, 351

\bibitem[{{Belczynski} {et~al.}(2012){Belczynski}, {Wiktorowicz}, {Fryer},
  {Holz}, \& {Kalogera}}]{belczynski+2012}
{Belczynski}, K., {Wiktorowicz}, G., {Fryer}, C.~L., {Holz}, D.~E., \&
  {Kalogera}, V. 2012, \apj, 757, 91

\bibitem[{{Boyajian} {et~al.}(2012){Boyajian}, {von Braun}, {van Belle},
  {McAlister}, {ten Brummelaar}, {Kane}, {Muirhead}, {Jones}, {White},
  {Schaefer}, {Ciardi}, {Henry}, {L{\'o}pez-Morales}, {Ridgway}, {Gies}, {Jao},
  {Rojas-Ayala}, {Parks}, {Sturmann}, {Sturmann}, {Turner}, {Farrington},
  {Goldfinger}, \& {Berger}}]{boyajian+2012}
{Boyajian}, T.~S., {von Braun}, K., {van Belle}, G., {et~al.} 2012, \apj, 757,
  112

\bibitem[{{Cantrell} {et~al.}(2008){Cantrell}, {Bailyn}, {McClintock}, \&
  {Orosz}}]{cantrell+2008}
{Cantrell}, A.~G., {Bailyn}, C.~D., {McClintock}, J.~E., \& {Orosz}, J.~A.
  2008, \apjl, 673, L159

\bibitem[{{Cantrell} {et~al.}(2010){Cantrell}, {Bailyn}, {Orosz}, {McClintock},
  {Remillard}, {Froning}, {Neilsen}, {Gelino}, \& {Gou}}]{cantrell+2010}
{Cantrell}, A.~G., {Bailyn}, C.~D., {Orosz}, J.~A., {et~al.} 2010, \apj, 710,
  1127

\bibitem[{{Cardelli} {et~al.}(1989){Cardelli}, {Clayton}, \&
  {Mathis}}]{cardelli+1989}
{Cardelli}, J.~A., {Clayton}, G.~C., \& {Mathis}, J.~S. 1989, \apj, 345, 245

\bibitem[{{Casares} \& {Jonker}(2014)}]{casares+2014}
{Casares}, J., \& {Jonker}, P.~G. 2014, \ssr, 183, 223

\bibitem[{{Casares} {et~al.}(1997){Casares}, {Mart{\'{\i}}n}, {Charles},
  {Molaro}, \& {Rebolo}}]{casares+1997}
{Casares}, J., {Mart{\'{\i}}n}, E.~L., {Charles}, P.~A., {Molaro}, P., \&
  {Rebolo}, R. 1997, \na, 1, 299

\bibitem[{{Charles} \& {Coe}(2006)}]{charles+2006}
{Charles}, P.~A., \& {Coe}, M.~J. 2006, {Optical, ultraviolet and infrared
  observations of X-ray binaries}, ed. W.~H.~G. {Lewin} \& M.~{van der Klis},
  215--265

\bibitem[{{Chen} {et~al.}(2015){Chen}, {Gou}, {McClintock}, {Steiner},
  {Wu}, {Xu}, {Orosz}, \& {Xiang}}]{chen+2015}
{Chen}, Z., {Gou}, L., {McClintock}, J.~E., {et~al.} 2015, \apj, submitted


\bibitem[{{Cheng} {et~al.}(1992){Cheng}, {Horne}, {Panagia}, {Shrader},
  {Gilmozzi}, {Paresce}, \& {Lund}}]{cheng+1992}
{Cheng}, F.~H., {Horne}, K., {Panagia}, N., {et~al.} 1992, \apj, 397, 664

\bibitem[{{Corral-Santana} {et~al.}(2011){Corral-Santana}, {Casares},
  {Shahbaz}, {Zurita}, {Mart{\'{\i}}nez-Pais}, \&
  {Rodr{\'{\i}}guez-Gil}}]{corral+2011}
{Corral-Santana}, J.~M., {Casares}, J., {Shahbaz}, T., {et~al.} 2011, \mnras,
  413, L15

\bibitem[{{Dubus} {et~al.}(2001){Dubus}, {Hameury}, \& {Lasota}}]{dubus+2001}
{Dubus}, G., {Hameury}, J.-M., \& {Lasota}, J.-P. 2001, \aap, 373, 251

\bibitem[{{Eggleton}(1983)}]{eggleton+1983}
{Eggleton}, P.~P. 1983, \apj, 268, 368

\bibitem[{{Esin} {et~al.}(1997){Esin}, {McClintock}, \& {Narayan}}]{esin+1997}
{Esin}, A.~A., {McClintock}, J.~E., \& {Narayan}, R. 1997, \apj, 489, 865

\bibitem[{{Farr} {et~al.}(2011){Farr}, {Sravan}, {Cantrell}, {Kreidberg},
  {Bailyn}, {Mandel}, \& {Kalogera}}]{farr+2011}
{Farr}, W.~M., {Sravan}, N., {Cantrell}, A., {et~al.} 2011, \apj, 741, 103

\bibitem[{{Fender} {et~al.}(2010){Fender}, {Gallo}, \& {Russell}}]{fender+2010}
{Fender}, R.~P., {Gallo}, E., \& {Russell}, D. 2010, \mnras, 406, 1425

\bibitem[{{Fender} {et~al.}(1999){Fender}, {Garrington}, {McKay}, {Muxlow},
  {Pooley}, {Spencer}, {Stirling}, \& {Waltman}}]{fender+1999}
{Fender}, R.~P., {Garrington}, S.~T., {McKay}, D.~J., {et~al.} 1999, \mnras,
  304, 865

\bibitem[{{Fragos} \& {McClintock}(2015)}]{fragos+2015}
{Fragos}, T., \& {McClintock}, J.~E. 2015, \apj, 800, 17

\bibitem[{{Fryer} \& {Kalogera}(2001)}]{fryer+2001}
{Fryer}, C.~L., \& {Kalogera}, V. 2001, \apj, 554, 548

\bibitem[{{Gelino}(2004)}]{gelino+2004}
{Gelino}, D.~M. 2004, in Revista Mexicana de Astronomia y Astrofisica, vol. 27,
  Vol.~20, Revista Mexicana de Astronomia y Astrofisica Conference Series, ed.
  G.~{Tovmassian} \& E.~{Sion}, 214--214

\bibitem[{{Gelino} \& {Harrison}(2003)}]{gelino+2003}
{Gelino}, D.~M., \& {Harrison}, T.~E. 2003, \apj, 599, 1254

\bibitem[{{Gelino} {et~al.}(2001){Gelino}, {Harrison}, \&
  {McNamara}}]{gelino+2001}
{Gelino}, D.~M., {Harrison}, T.~E., \& {McNamara}, B.~J. 2001, \aj, 122, 971

\bibitem[{{Greene} {et~al.}(2001){Greene}, {Bailyn}, \& {Orosz}}]{greene+2001}
{Greene}, J., {Bailyn}, C.~D., \& {Orosz}, J.~A. 2001, \apj, 554, 1290

\bibitem[{{Ioannou} {et~al.}(2004){Ioannou}, {Robinson}, {Welsh}, \&
  {Haswell}}]{ioannou+2004}
{Ioannou}, Z., {Robinson}, E.~L., {Welsh}, W.~F., \& {Haswell}, C.~A. 2004,
  \aj, 127, 481

\bibitem[{{Khargharia} {et~al.}(2010){Khargharia}, {Froning}, \&
  {Robinson}}]{khargharia+2010}
{Khargharia}, J., {Froning}, C.~S., \& {Robinson}, E.~L. 2010, \apj, 716, 1105

\bibitem[{{Khargharia} {et~al.}(2013){Khargharia}, {Froning}, {Robinson}, \&
  {Gelino}}]{khargharia+2013}
{Khargharia}, J., {Froning}, C.~S., {Robinson}, E.~L., \& {Gelino}, D.~M. 2013,
  \aj, 145, 21

\bibitem[{{King} {et~al.}(1996){King}, {Harrison}, \& {McNamara}}]{king+1996}
{King}, N.~L., {Harrison}, T.~E., \& {McNamara}, B.~J. 1996, \aj, 111, 1675

\bibitem[{{Kochanek}(2014)}]{kochanek+2014}
{Kochanek}, C.~S. 2014, \apj, 785, 28

\bibitem[{{Kreidberg} {et~al.}(2012){Kreidberg}, {Bailyn}, {Farr}, \&
  {Kalogera}}]{kreidberg+2012}
{Kreidberg}, L., {Bailyn}, C.~D., {Farr}, W.~M., \& {Kalogera}, V. 2012, \apj,
  757, 36

\bibitem[{{Lasota}(2001)}]{lasota+2001}
{Lasota}, J.-P. 2001, \nar, 45, 449

\bibitem[{{Marshall} {et~al.}(2008){Marshall}, {Burles}, {Thompson},
  {Shectman}, {Bigelow}, {Burley}, {Birk}, {Estrada}, {Jones}, {Smith},
  {Kowal}, {Castillo}, {Storts}, \& {Ortiz}}]{marshall+2008}
{Marshall}, J.~L., {Burles}, S., {Thompson}, I.~B., {et~al.} 2008, in Society
  of Photo-Optical Instrumentation Engineers (SPIE) Conference Series, Vol.
  7014, Society of Photo-Optical Instrumentation Engineers (SPIE) Conference
  Series, 54

\bibitem[{{McClintock} {et~al.}(2014){McClintock}, {Narayan}, \&
  {Steiner}}]{mcclintock+2014}
{McClintock}, J.~E., {Narayan}, R., \& {Steiner}, J.~F. 2014, \ssr, 183, 295

\bibitem[{{Morningstar} {et~al.}(2014){Morningstar}, {Miller}, {Reis}, \&
  {Ebisawa}}]{morningstar+2014}
{Morningstar}, W.~R., {Miller}, J.~M., {Reis}, R.~C., \& {Ebisawa}, K. 2014,
  \apjl, 784, L18

\bibitem[{{Narayan} \& {McClintock}(2005)}]{narayan+2005}
{Narayan}, R., \& {McClintock}, J.~E. 2005, \apj, 623, 1017

\bibitem[{{Narayan} \& {McClintock}(2012)}]{narayan+2012}
---. 2012, \mnras, 419, L69

\bibitem[{{Narayan} {et~al.}(2014){Narayan}, {McClintock}, \&
  {Tchekhovskoy}}]{narayan+2014}
{Narayan}, R., {McClintock}, J.~E., \& {Tchekhovskoy}, A. 2014, {Energy
  Extraction from Spinning Black Holes Via Relativistic Jets}, ed. J.~{Bi{\v
  c}{\'a}k} \& T.~{Ledvinka}, 523

\bibitem[{{Narayan} \& {Yi}(1994)}]{narayan+1994}
{Narayan}, R., \& {Yi}, I. 1994, \apjl, 428, L13

\bibitem[{{Orosz} \& {Bailyn}(1997)}]{orosz+1997}
{Orosz}, J.~A., \& {Bailyn}, C.~D. 1997, \apj, 477, 876

\bibitem[{{Orosz} {et~al.}(1996){Orosz}, {Bailyn}, {McClintock}, \&
  {Remillard}}]{orosz+1996}
{Orosz}, J.~A., {Bailyn}, C.~D., {McClintock}, J.~E., \& {Remillard}, R.~A.
  1996, \apj, 468, 380

\bibitem[{{Orosz} \& {Hauschildt}(2000)}]{orosz+2000}
{Orosz}, J.~A., \& {Hauschildt}, P.~H. 2000, \aap, 364, 265

\bibitem[{{Orosz} {et~al.}(2014){Orosz}, {Steiner}, {McClintock}, {Buxton},
  {Bailyn}, {Steeghs}, {Guberman}, \& {Torres}}]{orosz+2014}
{Orosz}, J.~A., {Steiner}, J.~F., {McClintock}, J.~E., {et~al.} 2014, \apj,
  794, 154

\bibitem[{{{\"O}zel} {et~al.}(2010){{\"O}zel}, {Psaltis}, {Narayan}, \&
  {McClintock}}]{ozel+2010}
{{\"O}zel}, F., {Psaltis}, D., {Narayan}, R., \& {McClintock}, J.~E. 2010,
  \apj, 725, 1918

\bibitem[{{Peris} {et~al.}(2015){Peris}, {Vrtilek}, {Steiner}, {Vrtilek}, {Wu},
  {McClintock}, {Longa-Pe{\~n}a}, {Steeghs}, {Callanan}, {Ho}, {Orosz}, \&
  {Reynolds}}]{peris+2015}
{Peris}, C.~S., {Vrtilek}, S.~D., {Steiner}, J.~F., {et~al.} 2015, \mnras, 449,
  1584

\bibitem[{{Reid} {et~al.}(2014){Reid}, {McClintock}, {Steiner}, {Steeghs},
  {Remillard}, {Dhawan}, \& {Narayan}}]{reid+2014}
{Reid}, M.~J., {McClintock}, J.~E., {Steiner}, J.~F., {et~al.} 2014, \apj, 796,
  2

\bibitem[{{Remillard} \& {McClintock}(2006)}]{remillard+2006}
{Remillard}, R.~A., \& {McClintock}, J.~E. 2006, \araa, 44, 49

\bibitem[{{Remillard} {et~al.}(1992){Remillard}, {McClintock}, \&
  {Bailyn}}]{remillard+1992}
{Remillard}, R.~A., {McClintock}, J.~E., \& {Bailyn}, C.~D. 1992, \apjl, 399,
  L145

\bibitem[{{Reynolds} {et~al.}(2007){Reynolds}, {Callanan}, \&
  {Filippenko}}]{reynolds+2007}
{Reynolds}, M.~T., {Callanan}, P.~J., \& {Filippenko}, A.~V. 2007, \mnras, 374,
  657

\bibitem[{{Reynolds} {et~al.}(2008){Reynolds}, {Callanan}, {Robinson}, \&
  {Froning}}]{reynolds+2008}
{Reynolds}, M.~T., {Callanan}, P.~J., {Robinson}, E.~L., \& {Froning}, C.~S.
  2008, \mnras, 387, 788

\bibitem[{{Russell} {et~al.}(2013){Russell}, {Gallo}, \&
  {Fender}}]{russell+2013}
{Russell}, D.~M., {Gallo}, E., \& {Fender}, R.~P. 2013, \mnras, 431, 405

\bibitem[{{Shahbaz} {et~al.}(1997){Shahbaz}, {Naylor}, \&
  {Charles}}]{shahbaz+1997}
{Shahbaz}, T., {Naylor}, T., \& {Charles}, P.~A. 1997, \mnras, 285, 607

\bibitem[{{Shrader} \& {Gonzalez-Riestra}(1993)}]{shrader+1993}
{Shrader}, C.~R., \& {Gonzalez-Riestra}, R. 1993, \aap, 276, 373

\bibitem[{{Steiner} \& {McClintock}(2012)}]{steiner+2012b}
{Steiner}, J.~F., \& {McClintock}, J.~E. 2012, \apj, 745, 136

\bibitem[{{Steiner} {et~al.}(2013){Steiner}, {McClintock}, \&
  {Narayan}}]{steiner+2013}
{Steiner}, J.~F., {McClintock}, J.~E., \& {Narayan}, R. 2013, \apj, 762, 104

\bibitem[{{Steiner} {et~al.}(2012){Steiner}, {McClintock}, \&
  {Reid}}]{steiner+2012a}
{Steiner}, J.~F., {McClintock}, J.~E., \& {Reid}, M.~J. 2012, \apjl, 745, L7

\bibitem[{{Sutaria} {et~al.}(2002){Sutaria}, {Kolb}, {Charles}, {Osborne},
  {Kuulkers}, {Casares}, {Harlaftis}, {Shahbaz}, {Still}, \&
  {Wheatley}}]{sutaria+2002}
{Sutaria}, F.~K., {Kolb}, U., {Charles}, P., {et~al.} 2002, \aap, 391, 993

\bibitem[{{van der Hooft} {et~al.}(1998){van der Hooft}, {Heemskerk},
  {Alberts}, \& {van Paradijs}}]{vanderhooft+1998}
{van der Hooft}, F., {Heemskerk}, M.~H.~M., {Alberts}, F., \& {van Paradijs},
  J. 1998, \aap, 329, 538

\bibitem[{{Wu} {et~al.}(2015){Wu}, {Orosz}, {McClintock}, {Steeghs},
  {Longa-Pe{\~n}a}, {Callanan}, {Gou}, {Ho}, {Jonker}, {Reynolds}, \&
  {Torres}}]{wu+2015b}
{Wu}, J., {Orosz}, J.~A., {McClintock}, J.~E., {et~al.} 2015, \apj, 806, 92

\end{thebibliography}



\end{document}